\documentclass{PoS}

\usepackage{amsfonts}
\usepackage{graphicx}
\usepackage{amsmath}
\usepackage{bm}

\newcommand*{\chpt}{\raise0.4ex\hbox{$\chi$}PT}

\newcommand{\Delstar}{\ensuremath{\Delta^{\raise0.18ex\hbox{${\scriptstyle *}$}}}}
\def\gtwid{{\,\raise.35ex\hbox{$>$\kern-.75em\lower1ex\hbox{$\sim$}}\,}}
\def\ltwid{{\,\raise.35ex\hbox{$<$\kern-.75em\lower1ex\hbox{$\sim$}}\,}}
\def\leftvec{{\raise1.5ex\hbox{$\leftarrow$}\kern-1.00em}}
\def\rightvec{{\raise1.5ex\hbox{$\rightarrow$}\kern-1.00em}}
\def\half{{\scriptstyle \raise.2ex\hbox{${1\over2}$}}}
\def\threehalves{{\scriptstyle \raise.15ex\hbox{${3\over2}$}}}
\def\third{{\scriptstyle \raise.15ex\hbox{${1\over3}$}}}
\def\third{{\scriptstyle \raise.15ex\hbox{${1\over3}$}}}
\def\twothirds{{\scriptstyle \raise.15ex\hbox{${2\over3}$}}}
\def\fourth{{\scriptstyle \raise.15ex\hbox{${1\over4}$}}}

\newcommand{\cM}{\ensuremath{\mathcal{M}}}

\newcommand{\cbar}{\ensuremath{\overline{c}}}

\newcommand*{\bea}{\begin{eqnarray}}
\newcommand*{\eea}{\end{eqnarray}}
\newcommand*{\be}{\begin{equation}}
\newcommand*{\ee}{\end{equation}}

\title{Lattice studies of hadrons with heavy flavors}

\ShortTitle{Heavy quarks}

\author{\speaker{C.\ Aubin}\\
        Department of Physics, College of William and Mary, 
        Williamsburg, VA, USA\\
        E-mail: \email{caaubin@wm.edu}}

\abstract{I will discuss recent developments in lattice studies of hadrons composed of heavy quarks. I will mostly cover topics which are at a state of direct comparison with experiment, but will also discuss new ideas and promising techniques to aid future studies of lattice heavy quark physics.}

\FullConference{The XXVII International Symposium on Lattice Field Theory - LAT2009\\
		 July 26-31 2009\\
		 Peking University, Beijing, China}

\begin{document}

\bibliographystyle{JHEP}

\section{Introduction}\label{sec:intro}

Heavy flavor physics invariably plays an important role in the lattice community. As with the light quark sector, quantities involving heavy quarks can be studied to both test the methodology used in lattice calculations as well as testing the physics of the Standard Model. Depending on the situation, if there is a discrepancy between experiment and a lattice result, one must be able to determine the source of this discrepancy, and whether or not it actually exists. Thus, whether one is testing the lattice techniques or the Standard Model, reliable comparison with experimental measurements is crucial.

Reliable comparison with experiment takes two forms. First, one must have good quantitative control over all systematic uncertainties which enter into a calculation, giving credence to the results. However, there is another criterion which is implied but rarely explicitly stated, even when stressed as important in other arenas, and that is agreement among different calculations of the same quantity using different formulations. This is an area where the light quark sector (especially in kaon physics, see Vittorio Lubicz's review at this conference \cite{LubiczPlenaryLat09}) has a tremendous advantage over the heavy quark sector. This becomes important in cases such as the ``$f_{D_s}$ puzzle,'' where a significant discrepancy between experiment and the lattice calculation has shown up, and for two years has not yet gone away. This discrepancy exists between the experimental average and a single lattice calculation, and has not be reproduced by another. 

For this review, I divide my discussion into three parts.\footnote{Obviously I cannot provide an exhaustive review of what has been reported in the field during this past year. I will mostly focus on a few select results, and only on those with at least two flavors of dynamical light quarks. I apologize to all those whose calculations I didn't have time to discuss.} I begin with a couple of examples of spectroscopy calculations which simultaneously allow lattice calculations to be tested as well as predicting experimental measurements. Specifically in Sec.~\ref{sec:spect}, I discuss one meson example (that of the $B_c^*$ mass) and the single-bottom baryon spectrum, where there has been some interest during the last year when comparing lattice and experimental results. Then I will discuss cases of calculations which are more focussed on testing the Standard Model. In Sec.~\ref{sec:B}, I will discuss current results on $B$-mixing and decays, which is at a stronger point than most as it has a larger number of groups which allow for decent cross-checks. In Sec.~\ref{sec:D}, I will discuss $D$ decays, mostly sticking to a discussion of the possible discrepancy between the lattice calculation and experimental measurement of $f_{D_s}$. In the final part, I discuss newer ideas and what lies in the future for heavy quark physics. In Sec.~\ref{sec:New}, I discuss new ideas for extracting semileptonic form factors for $D$ decays, and in Sec.~\ref{sec:nonlep}, I present some initial ideas for attempting extractions of nonleptonic $B$ decay matrix elements from a lattice simulation. I finish with some conclusions in Sec.~\ref{sec:conc}.

\section{Spectroscopy}\label{sec:spect}

Lattice studies of the particle spectrum can play two roles in understanding Nature. First, it is an invaluable tool to test lattice techniques by calculating particle masses that are known experimentally to see that the methodology used is sound. At the same time, it can be use to predict masses for states that have not yet been seen. One nice feature in spectrum calculations that involve heavy flavors is that often lattice calculations have been done at roughly the same time as the experimental measurements, so there are more cases of predictions, as opposed to the usual postdictions, of the particle spectrum.

\begin{figure}[t]
\begin{center}
\includegraphics[width=4in]{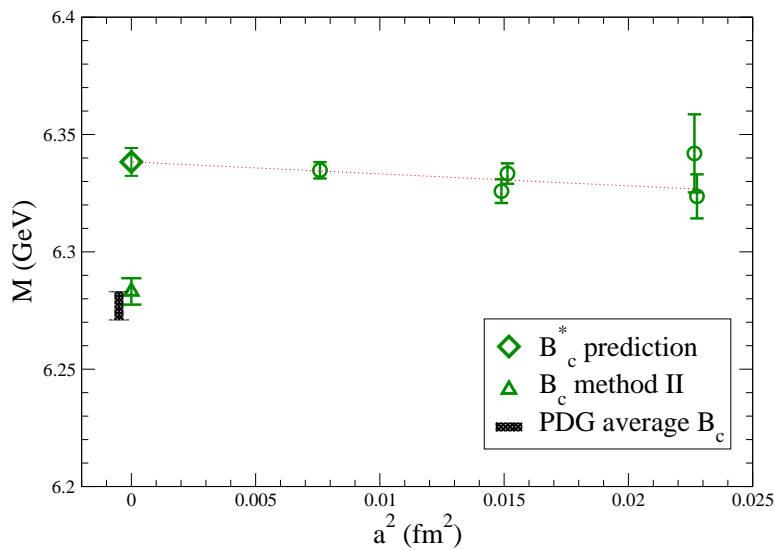}
\caption{Plot of $m_{B_c^*}$ versus the lattice spacing squared, showing the extrapolation to the continuum limit. On the same plot is the lattice determination of $m_{B_c}$ from \cite{Allison:2004be,Gregory:2008sk} compared with the PDG \cite{PDG}. Plot courtesy of E.~Gregory, from \cite{GregoryLat09}.}
\label{fig:BcStarPred}
\end{center}
\end{figure}

An example of this is the measurements of the $B_c$ system by the HPQCD collaboration. In \cite{Allison:2004be,Gregory:2008sk}, using NRQCD $b$ quarks and HISQ $c$ quarks in the valence sector and the MILC 2+1-flavor AsqTad configurations (three lattice spacings), HPQCD predicted the value of the $B_c$ mass, shortly afterwards to be confirmed by experiment \cite{Abulencia:2005usa}. This year, they have determined a prediction for the vector meson, the $B_c^*$ using the same techniques. In this case, they actually measure the ratio
\be
	\frac{m_{B_c^*} - m_{B_c}}{m_{B_s^*} - m_{B_s}}\ ,
\ee
and use the experimental value $(m_{B_s^*} - m_{B_s})^{\rm exp} = 49.0(1.5)$. The results are shown in Fig.~\ref{fig:BcStarPred}, and they obtain $m_{B_c^*} - m_{B_c} = 0.059(6)$ GeV for the mass splitting \cite{GregoryLat09}. 

On the baryon side, there have been many lattice results of baryons with one or more bottom quarks \cite{Lewis:2008fu,Burch:2008qx,Detmold:2008ww,Lin:2009rx,MeinelLat09}. Of these, there has been some interest in the last year in the single-bottom sector, due to an apparent discrepancy between lattice results and experiment. 

\begin{figure}[t]
\begin{center}
\includegraphics[width=4in]{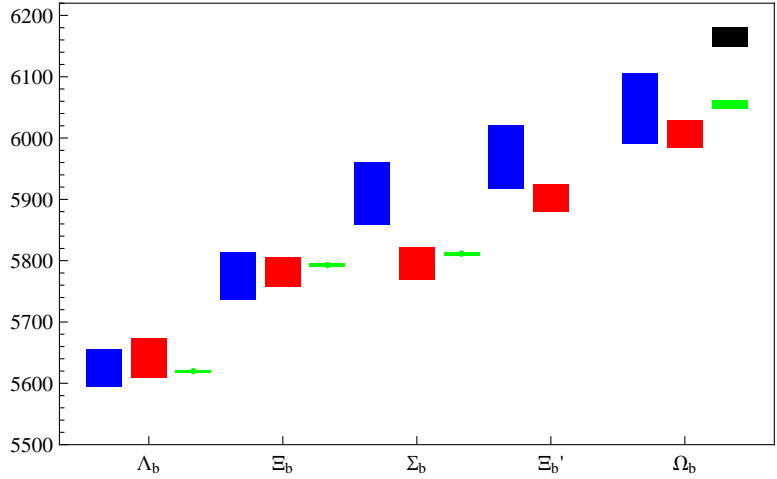}
\caption{Singly-bottom baryon spectrum from various lattice calculations and experiment. The data in red are using NRQCD heavy quarks \cite{Lewis:2008fu}, in blue use static heavy quarks \cite{Lin:2009rx}, and the green data are from experiment \cite{PDG,D0:2007ub,Aaltonen:2007rw,Aaltonen:2007un,D0:2008qm,Aaltonen:2009ny}. The measurement of the mass of the $\Omega_b$ differs between D0 (black) and CDF (green). Data from Refs.~\cite{Detmold:2008ww} (static heavy quarks) and \cite{MeinelLat09} (NRQCD quarks) are not shown for clarity, but is consistent with the other lattice calculations. The vertical scale is in MeV.}
\label{fig:staticlightspectrum}
\end{center}
\end{figure}

A summary of some results is shown in Fig.~\ref{fig:staticlightspectrum}, where I compare the lattice results\footnote{Note that only results which use 2+1-flavors of sea quarks are shown, and all results to date have only been presented at a single lattice spacing.} to the experimental results from the D0 collaboration \cite{D0:2007ub,D0:2008qm} and the CDF collaboration \cite{Aaltonen:2007rw,Aaltonen:2007un,Aaltonen:2009ny}. The puzzle as of last year was the agreement between the results shown in red (a calculation using NRQCD heavy quarks and Clover light quarks by Lewis and Woloshyn, \cite{Lewis:2008fu}) and D0 for all masses except for the $\Omega_b$, where there was an unexplained difference between the D0 measurement and theory. Later, various results using static heavy quarks \cite{Detmold:2008ww,Lin:2009rx} and NRQCD quarks \cite{MeinelLat09} confirmed the earlier lattice results, also shown in Fig.~\ref{fig:staticlightspectrum}.\footnote{The lattice results are consistent with theoretical understanding coming from Heavy Quark Effective Theory and various quark model pictures.}

The results from CDF published in May \cite{Aaltonen:2009ny} confirm the lattice picture, and thus the discrepancy has shifted from a lattice-experiment discrepancy to one between differing experiments, which I will not comment on further, as this is not within the scope of this review. What is important to note here is that a single lattice calculation which disagrees with experiment is not evidence enough for new physics, nor can we use this to claim either the experiment or the lattice calculation is flawed in some way. Rather, only when there are multiple lattice calculations (and experimental measurements) can we claim (dis)agreement between the the different sides.

\section{$B$-physics}\label{sec:B}

Moving on to calculations of matrix elements, I begin with the $b$-quark sector. Here, there have been calculations for the $B$ and $B_s$ meson decay constants as well as for the neutral $B_s$ and $B_d$ mixing parameters. Often these calculations are performed in concert with each other, because if one is interested in the $B$-parameters themselves (\emph{ie}, separated from the decay constants), one needs the decay constants.

As is well known, the decay constant $f_B$ governs the leptonic decay rate of a $B$-meson to a lepton-antineutrino pair. The $B$-parameter is defined by
$%\be\label{eq:B-param}
	\left\langle
	\overline{B}_q | O_{LL} | B_q
	\right\rangle
	=
	\frac{8}{3}f_{B_q}^2 m_{B_q} B_{B_q}\ ,
$%\ee
defined in this manner so that $B_{B_q}=1$ in the vacuum saturation approximation, and $O_{LL}$ is the left-left weak operator governing this transition. The $B_{B_q}$ parameter enters explicitly into the expression for the oscillation frequency of the $B_q$ meson, given by
\be\label{eq:deltamq}
	\Delta m_q = 
	\frac{G_F^2 m_W^2}{6\pi^2}
	\left|V_{tq}V^*_{tb}\right|^2 \eta_2^B S_0(x_t) m_{B_q}
	f_{B_q}^2 \hat B_{B_q}\ ,
\ee
where the left-hand side is determined experimentally, and only the decay constant and $B_{B_q}$ on the right-hand side are needed nonperturbatively. What is generally of most interest phenomenologically is the CKM matrix elements in Eq.~(\ref{eq:deltamq}), specifically the ratio
\be\label{eq:VtdoverVts}
	\frac{\left| V_{td} \right|}{\left| V_{ts} \right|}
	=
	\frac{f_{B_s}\sqrt{B_{B_s}} }{f_{B_d}\sqrt{B_{B_d}} }
	\sqrt{\frac{\Delta m_d}{\Delta m_s}
	\frac{m_{B_s}}{m_{B_d}}}\ .
\ee
The nice feature of this ratio is that only the quantities that are directly measured either on the lattice or experimentally are required to obtain this ratio. The quantity from the lattice is the ratio of the mixing parameters and decay constants, and is denoted by $\xi = (f_{B_s}\sqrt{B_{B_s}})/(f_{B_d}\sqrt{B_{B_d}})$. Additionally, as with any ratio, many systematic errors will cancel, and as such a more precise determination of $\xi$ can be made compared with any of the quantities individually. 

In Table~\ref{tab:Bmesonsims} I summarize the various collaborations who have calculated either the decay constants or the mixing parameters, or both (or in some cases just ratios of them). I list the lattice spacing(s) used, the minimum pion mass, as well as the formulations for the light and heavy quark actions. All simulations use the same valence and sea light quarks, except for that labelled ``HPQCD II,'' which uses HISQ valence light quarks and AsqTad sea quarks, and this is the main difference between the two HPQCD calculations listed. Additionally, while there are preliminary results for the HPQCD II calculation, no numbers have yet been presented resulting from this calculation. 

\begin{table}[t]
\begin{center}
\begin{tabular}{cccccccc}
\multicolumn{7}{c}{2+1 flavor}\\
\hline
Group & $a$ & 
$m_{\pi,\rm min}$ & $q$ & $Q$ & $f_B,f_{B_s}$ & $B_B$ & Refs \\
 & (fm) & (MeV) & &&& &\\
\hline
Fermilab/MILC & 0.09, 0.12 & 230 & Asqtad & Fermilab &\checkmark &\checkmark & \cite{Bernard:2009wr,KronfeldLat09}\\
HPQCD I & 0.09, 0.12 & 260 & Asqtad & NRQCD & \checkmark & \checkmark 
 & \cite{Gamiz:2009ku}\\
HPQCD II &0.09, 0.12, 0.15 & 320 & HISQ & NRQCD & & & \cite{JunkoLat09}\\
RBC/UKQCD & 0.11 & 400 & DW & Static & \checkmark & \checkmark& \cite{WitzelLat09}\\
\hline
\multicolumn{7}{c}{2 flavor}\\
\hline
Group & $a$ & 
$m_{\pi,\rm min}$ & $q$ & $Q$ & $f_B,f_{B_s}$ & $B_B$ & Refs\\
 & (fm) & (MeV) & &&&& \\
\hline
ETMC & 0.05, 0.065, & 300 & TM & Static/TM & \checkmark & & \cite{LubiczLat09}\\
& 0.085, 0.10 &&&&&&\\
Burch et al & 0.11,0.16 & 350 & CI & Static & \checkmark& & \cite{Burch:2008qx} \\
\end{tabular}
\end{center}
\caption{A summary of the various calculations of the $B,B_s$ meson decay constants and mixing parameters. $q$ denotes the light quark action (where DW = Domain-wall, TM = Twisted Mass Wilson, and CI = Chirally Invariant), and $Q$ denotes the heavy quark action used. A checkmark denotes that the quantity (or at least the ratio of quantities) have been reported before or during this conference.}
\label{tab:Bmesonsims}
\end{table}%

\begin{table}[t]
\begin{center}
\begin{tabular}{ccccc}
\multicolumn{5}{c}{2+1 flavor}\\
\hline
Group & $f_B$ (MeV) & $f_{B_s}$ (MeV) & $f_{B_s}/f_B$ & $\xi$\\
\hline
Fermilab/MILC & 195(11)${}^*$  & 243(11)${}^*$ &
$1.245(43)
%$[0.803(28)]^{-1}
{}^*$ 
 &  1.205(50) \\
HPQCD I & 190(13) & 231(15) & 1.226(26) &   1.258(33)  \\
RBC/UKQCD (APE)  & &  & 1.20(14) &  1.187(112)  \\
RBC/UKQCD (HYP2) & &  & 1.19(16) &  1.18(19)  \\
\hline
\multicolumn{5}{c}{2 flavor}\\
\hline
Group & $f_B$ & $f_{B_s}$& $f_{B_s}/f_B$ & $\xi$\\
\hline
ETMC & 203(17) & 247(16) & 1.22(6) &  \\
Burch \emph{et al} (0.16 fm) & & & 1.108(29) & \\
Burch \emph{et al} (0.11 fm)& & & 1.089(41) & \\
\end{tabular}
\caption{A summary of results from the various collaborations for the $B$ decay constants and mixing parameter. The asterisk on the Fermilab/MILC numbers denote that these are the results have not been updated since Lattice 2008. For the Burch \emph{et al} calculation, no continuum limit was taken for the 2-flavor results, so I list them separately.}
\label{tab:fB-B_results}
\end{center}
\end{table}%

The Fermilab/MILC collaboration has not updated their results for the $B$ decay constants this year, however they have for their determination of $B_B$, or more precisely, $\xi$. As discussed in the parallel session \cite{KronfeldLat09}, the group has checked their results for the perturbative renormalization, and as such has a more well-defined determination of the systematic errors. For the ratio $\xi$, this is a small percentage ($\sim0.2\%$), and not the dominant error. Most of the uncertainties tend to cancel in the ratio, and the only two errors that remain dominant are the light quark discretization and chiral fits (using staggered light quarks, they implement staggered \chpt\ fits \cite{Aubin:2003mg,Aubin:2005aq,BBschpt}), and the statistical, which are 2.8\% and 3.1\%, respectively. One can see in Table~\ref{tab:fB-B_results} that these two dominate the total error in $\xi$, which is roughly 4.3\%. 

The RBC/UKQCD calculation, being an initial study, has fewer data than the Fermilab/MILC group and only a single lattice spacing. However, while the formulations used are different, both the central values an sources of the largest systematic errors are similar. They use static quarks for the $b$ quark, using two different smearings (APE and HYP2) to control the heavy quark discretization errors, and domain wall light quarks for both the valence and sea. After performing the chiral fits, their errors too are dominated by statistics, discretization effects, and \chpt. Of the two smearings, they find smaller systematic errors coming from APE smearing, of a combined 5.4\%+7.7\% (stat+sys) on $\xi$ for the APE smearing compared with 4.0\%+15.1\% with HYP2 smearing. A similar (although not as great) reduction of errors is seen for the ratio of $f_{B_s}/f_B$, and both are shown in Table~\ref{tab:fB-B_results}.

As for the 2-flavor simulations, there exist currently only calculations of the decay constants, and not $\xi$. ETMC performed simulations using both static quarks and twisted mass Wilson quarks for the $b$ quark on four lattice spacings, controlling the continuum limit quite well. Their chiral fits are not using the relevant \chpt\ for tmWilson quarks, but instead are polynomials in the heavy-light mass, $m_{hq}$, and $1/m_{hq}$. As for the calculation of Burch \emph{et al}, their primary focus was on a determination of the excited hadron spectrum involving heavy quarks using a variational approach with chirally invariant light quarks and static heavy quarks. This allows one to obtain ratios of the decay constants quite easily, although they have not (for the dynamical case) performed a continuum extrapolation, and as such the systematics are incomplete. Both of these collaborations' results are shown in Table~\ref{tab:fB-B_results}. 

One can see in Table~\ref{tab:fB-B_results} that the results from the different collaborations are quite consistent for the various quantities. What is useful here is that there are various caculations of the same quantities, a couple of which having similarly sized (and well-controlled) uncertainties. Of course these quantities are also of tremendous importance to connect with experiment and especially to constrain CKM matrix elements \cite{RuthLat09}.

\begin{figure}[t]
\begin{center}
\includegraphics[width=4in,angle=270]{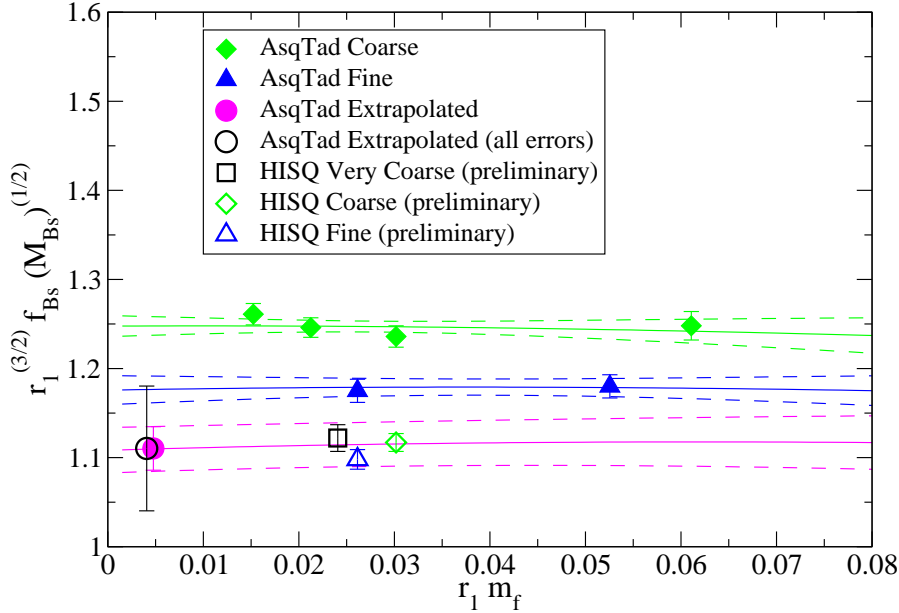}
\caption{Preliminary results from the HPQCD collaboration for $f_{B_s}\sqrt{m_{B_s}}$ using HISQ valence quarks, compared to using Asqtad. Figure courtesy of J.\ Shigemitsu, and presented in \cite{JunkoLat09}.}
\label{fig:HPQCDII}
\end{center}
\end{figure}

In addition, HPQCD has repeated their calculation of $B$-related quantities replacing the valence AsqTad quarks with HISQ quarks (still using NRQCD for the $b$ quark), and have presented preliminary results at this conference \cite{JunkoLat09}. As there are no final results yet, I will only show some indicative results for $f_{B_s}$, in Fig.~\ref{fig:HPQCDII}. This plot shows $f_{B_s}\sqrt{m_{B_s}}$ as a function of the valence quark mass for two (three) different lattice spacings for the AsqTad (HISQ) calculation, as well as the continuum extrapolated AsqTad results. One can see very clearly that difference in discretization errors between the AsqTad and HISQ calculations, which again shows how the HISQ action reduces the lattice spacing errors significantly. This is quite important as we have seen that lattice discretization errors tend to be one of the more dominant contributions to systematic uncertainties.

\section{$f_{D_s}$ and $|V_{cs}|$}\label{sec:D}

In the charm quark sector, there has been an ongoing ``puzzle'' for the last couple of years with regards to the $D_s$ decay constant. This puzzle was introduced when the HPQCD collaboration published an extremely precise result ($\sim 1.2\%$) for $f_{D_s}$ \cite{Follana:2007uv} that disagreed by roughly 3$\sigma$ from experimental results \cite{Belle:2007ws,CLEO:2009th,CLEO:2009ux}. This disagreement is enhanced by the fact that all other quantities calculated using the same methodology [HISQ valence quarks ($u,d,s,c$) on AsqTad sea quarks] agree quite well with the experimental measurements. 

To discuss this possible signal for new physics, I first outline the ingredients. The decay rate for $D_s\to \ell\bar\nu$ is given by
\be\label{eq:Ds_decay}
	\Gamma(D_s\to \ell\bar\nu)
	=
	\frac{G_F^2 m_\ell^2 m_{D_s}}{8\pi}
	\left(1 - \frac{m_\ell^2}{m_{D_s}^2}\right)^2
	f_{D_s}^2 |V_{cs}|^2\ ,
\ee
with the lattice being able to provide $f_{D_s}$, and experimental measurements of the decay rate can determine only the product $f_{D_s} |V_{cs}|$ by measuring $\Gamma$. Assuming the CKM matrix is unitary, we have the relation $|V_{ud}| = |V_{cs}| + \mathcal{O}(\lambda^4)$, where $\lambda^4\approx0.003$ and $|V_{ud}|$ is known extremely well, so one can use this substitution with the experimental measurement to extract $f_{D_s}$. 

As of last year's conference, the discrepancy remained (see E.~G\'amiz's review from Lattice 2008 \cite{Gamiz:2008iv}). The results were\footnote{I listed only the averages of the experimental results which measure absolute branching ratios for this decay, which is not what is listed in the Particle Data Book. The reason for this is to remove the extra (possibly large) systematic coming from the relative branching ratio measurements.}
\bea
	{\rm HPQCD}:\ f_{D_s} & = & (241\pm3)\, {\rm MeV}\nonumber\\
	{\rm CLEO+Belle}:\ f_{D_s} & = & (270.4\pm7.3\pm3.7)\, {\rm MeV}
	\ .\nonumber
\eea
Here I do not list the value determined by the Fermilab/MILC collaborations \cite{Bernard:2009wr}, which was consistent with HPQCD with a factor of three larger total uncertainty (and as such, it was also consistent with experiment). 

During the last year, there has been significant progress both on the theoretical side as well as experimental. CLEO-c has published new results, which I show in Table \ref{tab:fDs}. In this table, adapted from Ref.~\cite{CLEO:2009ux}, I show the most recent results from Belle and CLEO-c which determine the absolute branching ratios for $D_s^+\to\mu^+\nu$ and $D_s^+\to\tau^+\nu$.\footnote{Ref.~\cite{CLEO:2009ux} includes results that measure branching ratios relative to $D_s\to\phi\pi$. Again, I do not consider these due to the large systematic uncertainties that plague these measurements.}

One can see that the average of the CLEO-c and Belle numbers has come down significantly, to
\[
	f_{D_s} = 261.2\pm 6.9\ {\rm MeV}\ .
\]
The error here is also reduced some, so the discrepancy between this and HPQCD is around 2.5$\sigma$ now, which is reduced but possibly significant. 

\begin{table}[tb]
\begin{center}

\caption{Summary of experimental results for ${\cal{B}}(D_s^+\to \mu^+\nu)$,
${\cal{B}}(D_s^+\to \tau^+\nu)$, and $f_{D_s^+}$, taking into account results from absolute branching fractions only. Results have been updated for the new value of the $D_s$ lifetime of 0.5 ps \cite{PDG}. (This table adapted from Table V of Ref.~\cite{CLEO:2009ux}.) \label{tab:fDs}}\label{tab:cleo_fDs}
\begin{tabular}{llcc}\hline\hline
Exp. & Mode  & ${\cal{B}}$& $f_{D_s^+}$ (MeV) \\
\hline
Belle \cite{Belle:2007ws}  & $\mu^+\nu$ & $(6.38\pm 0.76\pm 0.52)\cdot 10^{-3}$& $274\pm 16 \pm 12 $ \\
CLEO-c  \cite{CLEO:2009th}& $\tau^+\nu$ & $(5.30\pm 0.47\pm 0.22)\cdot 10^{-2}$& $252.5\pm 11.1 \pm 5.2$ \\
CLEO-c \cite{CLEO:2009ux}& $\mu^+\nu$ & $(5.65\pm 0.45\pm 0.17)\cdot 10^{-3}$ & $257.3\pm 10.3\pm 3.9$\\
CLEO-c \cite{CLEO:2009ux}& $\tau^+\nu$ & $(6.42\pm 0.81\pm 0.18)\cdot 10^{-2}$& $278.7\pm 17.1 \pm 3.8 $ \\
%CLEO-c & \multicolumn{2}{c}{combined above 2 results using SM} & & $263.3\pm 8.2\pm 3.9$  \\
CLEO-c & \multicolumn{2}{c}{combined all CLEO-c results}   &$259.5\pm 6.6\pm 3.1$  \\
\hline
\multicolumn{3}{l}{Average of CLEO and Belle results above, radiatively corrected}  & $261.2\pm 6.9$\\
 \hline
\end{tabular}
\end{center}

\end{table}

\begin{figure}[t]
\begin{tabular}{ccc}
\includegraphics[width=2.5in]{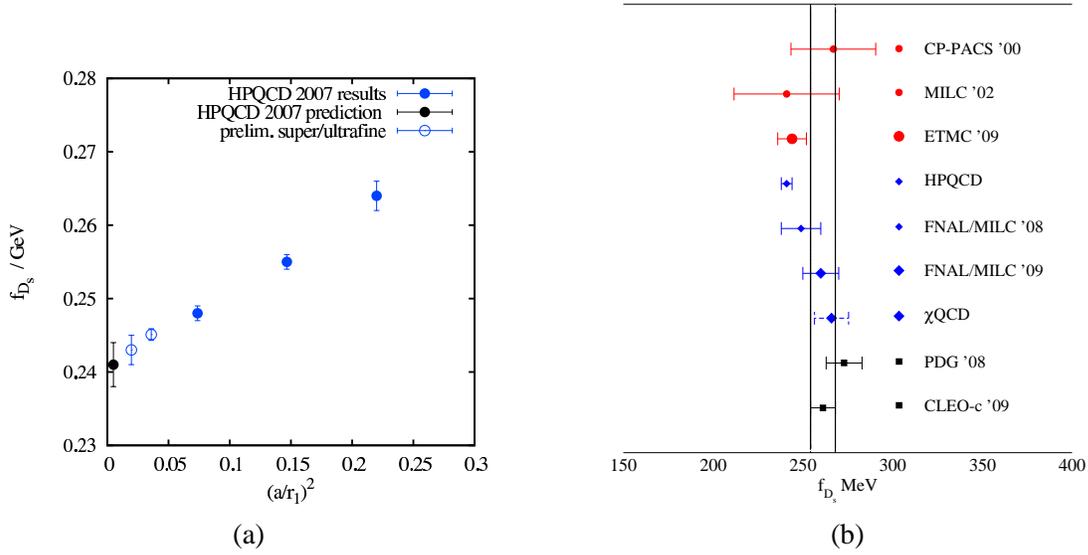}
%\caption{New HPQCD results}
%\label{fig:HPQCDfDs}
&$\qquad$&
\includegraphics[width=2.5in]{fDsPlot.eps}\\
(a) &&(b)
\end{tabular}
\caption{
(a)Updated HPQCD results including the superfine and ultrafine MILC lattices. The black point is the extrapolation including only the three original data points (filed circles).
(b) All results with either 2 (red) or 2+1 (blue) flavors of dynamical quarks for $f_{D_s}$, compared with both the PDG value as well as the most recent CLEO-c result. The larger symbols (for ETMC, Fermilab/MILC '09, and $\chi$QCD) denote lattice results presented within the last year. The dashed error bars on the $\chi$QCD result indicate that these errors include only statistical uncertainties.}\label{fig:fDs}
%\end{center}
\end{figure}

HPQCD has performed more simulations, adding two lattice spacings (using the superfine and ultrafine MILC lattices, $a\approx 0.06$ fm and $a\approx0.045$ fm, respectively). The analysis has not been complete, but the new data points are shown in Fig.~\ref{fig:fDs}(a), with the previous three lattice spacings as well as the 2007 result (which only involves a chiral/continuum extrapolation from the heavier three lattice spacings, not the new data). One can see that the new points fall right in line with the other three, and as such at first glance this will not change the final central value much, only a reduction of errors will emerge.

However, one thing that has not been included is a new determination of the relative scale, $r_1$. Since $r_1$ enters the calculation in a highly non-trivial way, it is difficult to know how a new determination will affect the final results. As such, it is hard to predict what the updated value for $f_{D_s}$ will be with the new HPQCD data and inclusion of the the new $r_1$ value.

In addition, new preliminary results have been presented at this conference from the Fermilab/MILC \cite{SimoneLat09} and $\chi$QCD \cite{KehfeiLat09} groups, as well as a 2-flavor result from ETMC, which was published in Ref.~\cite{Blossier:2009bx}. The Fermilab/MILC result is an update of previous results using AsqTad light quarks and Fermilab heavy quarks on three lattice spacings. The $\chi$QCD calculation uses the Overlap formulation for both the light and charm valence quarks on the RBC/UKQCD Domain-Wall Fermion lattices. This calculation is done on two lattice spacings, with only fifty configurations. The ETMC calculation uses twisted-mass quarks for charm quark as well as the light quarks and they have performed the simulation on three lattice spacings, but only have two dynamical quark flavors.

I show the various 2 and 2+1-flavor results to date in Fig.~\ref{fig:fDs}(b). The latest 2-flavor result from ETMC \cite{Blossier:2009bx} agrees with the previous results from several years ago \cite{AliKhan:2000eg,Bernard:2002pc}, with significantly improved errors. The two Fermilab/MILC numbers are from those reported at Lattice 2008 \cite{Bernard:2009wr} and a newer number presented at this conference \cite{SimoneLat09}. One thing to note is that except for the $\chi$QCD calculation \cite{KehfeiLat09}, all numbers shown on the plot have included both statistical and systematic uncertainties (added in quadrature), while $\chi$QCD currently only quotes a statistical error. Another interesting point is that the new Fermilab/MILC number has shifted upwards, slightly outside (but not significantly) the errors of the HPQCD calculation. 

Before making any conclusions, however, I would like to point out that CKM unitarity plays a vital role in the experimental determination, and the question remains as to whether or not the relation $|V_{ud}|\approx|V_{cs}|$ actually holds, and the lattice can play a part in understanding this. This involves calculating the semileptonic form factors for $D_s\to K\ell\bar\nu$ to determine $|V_{cs}|$, which has been done for 2+1-flavors by the Fermilab/MILC group (initially presented in Ref.~\cite{Aubin:2004ej}), with the most recent determination presented last year being
\be\label{eq:VcsLat}
	|V_{cs}| = 1.015\pm 0.015\pm0.106\ .
\ee
This is of course consistent with unitarity, but with an 11\% error, this is not a stringent constraint. New/updated calculations by ETMC \cite{SimulaLat09}, Fermilab/MILC \cite{SimoneLat09}, and the Regensburg group \cite{EvansLat09} have been discussed at this conference, but no results for the CKM matrix element has yet been presented.

So the question remains, is there really an $f_{D_s}$ puzzle? There is no conclusive evidence that there is a discrepancy between the lattice calculations and experiment as yet. This emphasizes the need for other determinations of $f_{D_s}$ to match the precision quoted by HPQCD. As it stands now, it may be that the HPQCD is an outlier, and this discrepancy may slowly disappear (as these ``new physics'' indications often do). However, with more calculations improving, this could be an indication of new physics in the charm quark sector. 

\section{New Methods for semileptonic decays}\label{sec:New}

As discussed at the end of the previous section, there is an increasing need for precision in calculations of semileptonic decays of $D\to P\ell\nu$, where $D$ is either a $D^+$ or $D_s$ meson, and $P$ can be $\pi$ or $K$. The difficulty lies in the extraction of the form factors from the three-point functions $\left\langle P | V^\mu | K\right\rangle$ as a function of $q^2$. One limiting factor is the cost of the calculation as one wishes to change the momentum transfer $q$, as generally this requires an additional propagator generation for each $q$. Thus, to get both the shape and normalization of the form factors as a function of $q^2$ this can become an expensive calculation. 

One solution is to minimize the number of Dirac operator inversions that are required at the cost of introducing noise. Such stochastic methods are not new (for example, see Refs.~\cite{Foley:2005ac,Peardon:2009gh}), and when implemented can easily save on the cost of the simulation. In particular, this was discussed and preliminary results were shown for the shape of the form factor at this conference \cite{EvansLat09}, and this is an interesting approach to the problem.

Another approach which can lead to more immediate phenomenological impact has been presented by H.\ Na \cite{NaLat09} for the HPQCD collaboration at this conference. Looking at the expression for the differential decay rate, we have
\be
	\frac{d\Gamma(D\to K\ell\nu)}{dq^2}
	=
	\frac{G_F^2|\mathbf{p}_K|^3}{24\pi^3}|f_+^{D\to K}(q^2)|^2
	|V_{cs}|^2\ ,
\ee
where $\mathbf{p}_K$ is the three-momentum of the outgoing kaon. What is then needed from the lattice is the particular form factor $f_+$, which comes from the parametrization
\be\label{eq:Dff}
	\left\langle K|\bar s \gamma^\mu c | D_s\right\rangle
	=
	\left[
	p_D^\mu + p_K^\mu - \frac{m_D^2 - m_K^2}{q^2}q^\mu
	\right]
	f_+(q^2)
	+\frac{m_D^2 - m_K^2}{q^2}q^\mu f_0(q^2)	\ .
\ee
However, what is generally extracted from experimental calculations is the normalization of the form factors, or more specifically, $|f_+(0)||V_{cs}|$. Thus, a lattice determination of $f_+(0)$ is sufficient to determine $|V_{cs}|$.

This can be acheived with two simple ingredients. First, there is a kinematic constraint that states that the two form factors in Eq.~(\ref{eq:Dff}) are equal at $q^2=0$: $f_+(0) = f_0(0)$. Second, that the scalar form factor can be related to the vector form factor by
\be
	q_\mu \left\langle K|\bar s \gamma^\mu c | D_s\right\rangle
	=
	\left\langle K|\bar s  c | D_s\right\rangle\ , 
\ee
which leads to the important relation
\be
	f_+(0) = f_0(0) = \frac{m_c - m_s}{m_D^2 - m_\pi^2}
	\left(\left\langle K|\bar s c | 
	D_s\right\rangle\right)_{q^2 = 0}\ ,
\ee
and the right hand side can be extracted with less noise than with the more direct method. Preliminary results, tested by looking at $D_s\to \eta_s\ell\nu$ decays, were presented at this conference \cite{NaLat09}, and show a promising method to calculate the form factor at $q^2=0$ with few percent precision. This is an essential goal for any method to calculate $|V_{cs}|$, or any other CKM matrix element.

\section{Nonleptonic decays}\label{sec:nonlep}

When looking at decays of heavy-flavored states on the lattice, the primary focus is that of leptonic or semileptonic decays. The reason for this is that in these types of decays, there is either one or zero hadrons in the final state. One reason for this is simplicity: As you add hadrons in the final state, the calculation becomes more complicated and noisy. As such, $f_D$ (zero hadrons in the final state) is much more simple to calculate than the form factors for $D\to\pi\ell\nu$, which has a single hadron in the final state. Of course, there is another reason for not looking at processes with more than one hadron in the final state, and that is because it is not directly accessible. There is the Maiani-Testa ``no-go'' theorem that states that physical Minkowskian amplitudes are not obtainable from Euclidean correlation functions in an infinite volume \cite{Maiani:1990ca}.

However, this is an obvious limitation in the field, as there is a wealth of information hidden within nonleptonic decays of hadrons. A large amount of effort, for example, has been spent trying to extract the amplitudes for $K\to2\pi$ decays (see for example Refs.~\cite{Lellouch:2000pv,Kim:2002np,Lin:2003tn,Christ:2005gi,Kim:2008dr,LightmanLat09,ChristLat09}), even though it is limited by the no-go theorem. These various references apply various tricks and other techniques to bypass the Maiani-Testa theorem in order to attempt a lattice calculation of $K\to2\pi$.

In fact, the decay of a kaon to two pions is complicated by several other features besides the Maiani-Testa theorem. It can occur in two channels, the $\Delta I = 3/2$ channel and the $\Delta I = 1/2$ channel. The former is more straightforward, but the latter is complicated by the addition of disconnected diagrams and lower-dimensional operator mixing (leading to power divergences as the lattice spacing goes to zero), among other things. However, it is such an important quantity that there is an industry trying to circumvent these difficulties, so as to make any progress on a lattice determination of the $K\to2\pi$ amplitudes.

In the same vein, there is a lot to be gained from studying nonleptonic decays of heavy-light mesons. For concreteness, I will focus on $B\to D \pi$ or $B\to DK$ decays, and what can be gained from understanding these amplitudes \cite{AubinSoni}. The reason for this is primarily one of simplicity: These amplitudes, as I will discuss shortly, are free from several difficulties found in other processes, and they give essential insight into the CKM unitarity triangle. 

The effective Hamiltonian which governs these decays is given by \cite{Buras:1991jm}
\be\label{eq:Heff}
	H_{\rm eff} = 
	\frac{G_F}{\sqrt{2}} \sum_{j=1,2}\sum_{i=d,s} [
	V_{cb}^*V_{ui} C_j(\mu) Q^{b\to c,i}_j 
	+ V_{ub}^*V_{ci} C_j(\mu) Q^{b\to \cbar,i}_j
	+ {\rm h.c.}]\ .
\ee	
The operators are denoted with $j=1,2$ for the color mixed and color unmixed four-quark operators. What is most interesting is the ratio between the amplitudes for $B\to DK$ and $B\to \overline{D}K$ (or similarly with $K\to\pi$). The ratio \cite{PDG}
\be
	r_B = \left|\frac{\cM(B^- \to \overline{D}^0 K^-)}
	{\cM(B^- \to D^0 K^-)}\right|
	\sim 0.1-0.2
%	= \left|\frac{V_{ub}^*V_{cs}}{V_{cb}^*V_{us}}\right|
%	\times \left({\rm Pert.\ Th.}
%	\right)\times \left({\rm lattice}\right)\ ,
\ee
allows us to extract the CKM angle $\gamma$ \cite{Gronau:1990ra,Gronau:1994bn}, if the various matrix elements [from four of the operators in Eq.~(\ref{eq:Heff})] can be determined from lattice simulations. 

These matrix elements are actually much simpler to calculate than those in $K\to2\pi$ decays. For one, there are no penguin contributions or disconnected pieces. In fact, of the four possible $K\to2\pi$ diagrams (in the $\Delta I=1/2$ case), only one enters in this analogous $B\to DK$ decay. However, this does not remove the restriction imposed upon us by the Maiani-Testa theorem, and thus we must attempt to circumvent this problem.

An initial, if rather crude, approximation is to look to Chiral Perturbation Theory (\chpt), as has been done in the kaon case \cite{Bernard:1985wf}, and Heavy Quark Effective Theory (HQET) \cite{Manohar:2000dt}. If one works out the heavy-light \chpt, it is easy to show that at leading order in both $m_\pi$ and $1/m_{b,c}$:
\be\label{eq:chptRelation}
	(r_B)_{{\rm LO}\chi{\rm PT}} = \left|\frac{\cM(B_s \to \overline{D}^0)}
	{\cM(B_s \to D^0)}\right|\ .
\ee
In this relation, I have used the fact that at leading order in HL\chpt, both the $b$ and $c$ quarks are static, and thus the ``decays'' shown in Eq.~(\ref{eq:chptRelation}) are allowed via the weak operators in Eq.~(\ref{eq:Heff}). A similar (and theoretically more accurate) relation can be found for the $B\to D\pi$ matrix elements. 

While valid at leading order, there could easily be large corrections coming from higher order effects. The corrections coming from HQET, where $m_b$ and $m_c$ are not infinite (but still degenerate) are most likely mild, as these corrections often can be absorbed into redefinitions of the various couplings \cite{Boyd:1994pa}. The two corrections that are not likely to be mild are those coming from $m_b\ne m_c$, and those coming from higher orders in the pion/kaon mass.

For the first correction, where the heavy quarks are not degenerate, one can see immediately where this will be a problem. Setting the heavy quark masses to their physical values, the outgoing kaon or pion will carry a large momentum, well above the regime where \chpt\ is reliable. In this case, one would not expect the determination of $r_B$ in Eq.~(\ref{eq:chptRelation}) to be trustworthy. However, in the regime where $m_b\sim m_c$, this is a reasonable approximation.

As for higher orders in the light sector, often corrections involving one-loop diagrams with kaons can approach the 15-30\% level, depending on the quantity. While this is an unacceptable uncertainty for determinations of CKM matrix elements, one could hope that much of this uncertainty is cancelled in the ratio that is desired. Without doing the explicit calculation (both of the \chpt\ and the lattice calculation), nothing can be ascertained about this uncertainty, but it must be kept in mind before making quantitative claims.

However, I would like to point out that this should not be taken as the ultimate approach to extract $\gamma$ and be able to make definitive claims with regards to the Standard Model. This is meant as a starting point for this particular set of quantities. In a particular limit ($m_b=m_c=\infty, m_{u,d}=m_s=0$), this relation is \emph{exact}, and allows for an initial approach to this problem from a lattice perspective. 

Ultimately, of course, methods to tackle the four-point functions themselves for $B\to DK$ and $B\to D\pi$ decays must be developed. These could include, but are not limited to, different (unphysical) kinematics to bypass the Maiani-Testa theorem or varying the boundary conditions (both of which have been used in the kaon sector). In addition to this, one must be able to get a handle on the final state interactions which are likely to be an important ingredient in these decays.

\section{Conclusions}\label{sec:conc}

While in recent years much progress has been made in heavy flavor physics on the lattice, there is still quite a bit that needs to be done. As in the light quark sector, there needs to be numerous cross checks between different calculations using different techniques to ensure credibility in lattice calculations as a whole. At the same time, there have been steps made in new methods to better constrain quantities that have already been determined. Finally, I have proposed initial steps in lattice calculations of nonleptonic decays of heavy-light states, which are essential for a complete lattice picture of the heavy quark sector.

\section*{Acknowledgements}

I would like to thank quite a few people who helped in my preparation for this review. Of those whose data I explicitly used, I would like to thank Tommy Burch, Christine Davies, R.~Todd Evans, Elvira G\'amiz, Eric Gregory, Andreas Kronfeld, Keh-Fei Liu, Stefan Meinel, Heechang Na, Kostas Orginos, Giancarlo Rossi, Junko Shigemitsu, Jim Simone, Silvano Simula, Amarjit Soni, Cecilia Tarantino, and Oliver Witzel. Additionally I appreciate the information sent to me by many others which I unfortunately did not have time to cover during my talk. Finally I would like to thank the organizers for inviting me to give this review, and for planning a very enjoyable meeting. This work was partially supported by the US Department of Energy, under contract no. DE-FG02-04ER41302.

\bibliography{refs}

\providecommand{\href}[2]{#2}\begingroup\raggedright\begin{thebibliography}{10}

\bibitem{LubiczPlenaryLat09}
V.~Lubicz,
\newblock plenary talk at this conference.

\bibitem{Allison:2004be}
{\bf HPQCD} Collaboration, I.~F. Allison {\em et~al.}, {\em Phys. Rev. Lett.}
  {\bf 94} (2005) 172001.

\bibitem{Gregory:2008sk}
E.~B. Gregory {\em et~al.}, {{\tt arXiv:0810.1845 [hep-lat]}}.

\bibitem{Abulencia:2005usa}
{\bf CDF} Collaboration, A.~Abulencia {\em et~al.}, {\em Phys. Rev. Lett.} {\bf
  96} (2006) 082002.

\bibitem{GregoryLat09}
E.~B. Gregory,
\newblock talk presented at this conference.

\bibitem{PDG}
{\bf Particle Data Group} Collaboration, C.~Amsler {\em et~al.}, {\em Phys.
  Lett.} {\bf B667} (2008).

\bibitem{Lewis:2008fu}
R.~Lewis and R.~M. Woloshyn, {\em Phys. Rev.} {\bf D79} (2009) 014502.

\bibitem{Burch:2008qx}
T.~Burch, C.~Hagen, C.~B. Lang, M.~Limmer, and A.~Schafer, {\em Phys. Rev.} {\bf D79} (2009) 014504.

\bibitem{Detmold:2008ww}
W.~Detmold, C.~J.~D. Lin, and M.~Wingate, {\em Nucl. Phys.} {\bf B818} (2009) 17.

\bibitem{Lin:2009rx}
H.-W. Lin, S.~D. Cohen, N.~Mathur, and K.~Orginos,
  {{\tt arXiv:0905.4120 [hep-lat]}}.

\bibitem{MeinelLat09}
  S.~Meinel, W.~Detmold, C.~J.~Lin and M.~Wingate,
  %``Bottom hadrons from lattice QCD with domain wall and NRQCD fermions,''
  arXiv:0909.3837 [hep-lat].
  %%CITATION = ARXIV:0909.3837;%%

\bibitem{D0:2007ub}
{\bf D0} Collaboration, V.~M. Abazov {\em et~al.}, {\em Phys. Rev. Lett.} {\bf
  99} (2007) 052001.

\bibitem{D0:2008qm}
{\bf D0} Collaboration, V.~M. Abazov {\em et~al.}, {\em Phys. Rev. Lett.} {\bf
  101} (2008) 232002.

\bibitem{Aaltonen:2007rw}
{\bf CDF} Collaboration, T.~Aaltonen {\em et~al.}, {\em Phys. Rev. Lett.} {\bf
  99} (2007) 202001.

\bibitem{Aaltonen:2007un}
{\bf CDF} Collaboration, T.~Aaltonen {\em et~al.}, {\em Phys. Rev. Lett.} {\bf
  99} (2007) 052002.

\bibitem{Aaltonen:2009ny}
{\bf CDF} Collaboration, T.~Aaltonen {\em et~al.},
  {{\tt arXiv:0905.3123 [hep-ex]}}.

\bibitem{Bernard:2009wr}
C.~Bernard {\em et~al.}, {\em PoS} {\bf LATTICE2008} (2008) 278,
  [{{\tt arXiv:0904.1895 [hep-lat]}}].

\bibitem{KronfeldLat09}
A.~Kronfeld,
\newblock talk presented at this conference.

\bibitem{Gamiz:2009ku}
{\bf HPQCD} Collaboration, E.~Gamiz, C.~T.~H. Davies, G.~P. Lepage,
  J.~Shigemitsu, and M.~Wingate, {\em Phys. Rev.} {\bf D80} (2009) 014503.

\bibitem{JunkoLat09}
J.~Shigemitsu,
\newblock talk presented at this conference.

\bibitem{WitzelLat09}
O.~Witzel,
\newblock talk presented at this conference.

\bibitem{LubiczLat09}
V.~Lubicz,
\newblock poster presented at this conference.

\bibitem{Aubin:2003mg}
C.~Aubin and C.~Bernard, {\em Phys. Rev.} {\bf D68} (2003) 034014.

\bibitem{Aubin:2005aq}
C.~Aubin and C.~Bernard, {\em Phys. Rev.} {\bf D73} (2006) 014515.

\bibitem{BBschpt}
C.~Bernard, L.~Laiho, and R.S.~Van~de Water,
\newblock unpublished.

\bibitem{RuthLat09}
R.S.~Van~de Water,
\newblock plenary talk at this conference.

\bibitem{Follana:2007uv}
{\bf HPQCD} Collaboration, E.~Follana, C.~T.~H. Davies, G.~P. Lepage, and
  J.~Shigemitsu, {\em Phys. Rev. Lett.} {\bf 100} (2008) 062002.

\bibitem{Belle:2007ws}
{\bf Belle} Collaboration, K.~Abe {\em et~al.}, {\em Phys. Rev. Lett.} {\bf 100}
  (2008) 241801.

\bibitem{CLEO:2009th}
{\bf CLEO} Collaboration, P.~U.~E. Onyisi {\em et~al.}, {\em Phys. Rev.} {\bf
  D79} (2009) 052002.

\bibitem{CLEO:2009ux}
{\bf CLEO} Collaboration, J.~P. Alexander {\em et~al.}, {\em Phys. Rev.} {\bf
  D79} (2009) 052001.

\bibitem{Gamiz:2008iv}
E.~Gamiz, {{\tt arXiv:0811.4146 [hep-lat]}}.

\bibitem{SimoneLat09}
J.~Simone,
\newblock talk presented at this conference.

\bibitem{KehfeiLat09}
K.-F. Liu,
\newblock talk presented at this conference.

\bibitem{Blossier:2009bx}
B.~Blossier {\em et~al.}, {\em JHEP} {\bf 07} (2009) 043.

\bibitem{AliKhan:2000eg}
{\bf CP-PACS} Collaboration, A.~Ali~Khan {\em et~al.}, {\em Phys. Rev.} {\bf
  D64} (2001) 034505.

\bibitem{Bernard:2002pc}
{\bf MILC} Collaboration, C.~Bernard {\em et~al.}, {\em Phys. Rev.} {\bf D66}
  (2002) 094501.

\bibitem{Aubin:2004ej}
{\bf MILC \& Fermilab Lattice} Collaboration, C.~Aubin {\em et~al.}, {\em Phys.
  Rev. Lett.} {\bf 94} (2005) 011601.

\bibitem{SimulaLat09}
S.~Simula,
\newblock talk presented at this conference.

\bibitem{EvansLat09}
R.~Evans,
\newblock talk presented at this conference.

\bibitem{Foley:2005ac}
J.~Foley {\em et~al.}, {\em Comput. Phys. Commun.} {\bf 172} (2005) 145.

\bibitem{Peardon:2009gh}
{\bf Hadron Spectrum} Collaboration, M.~Peardon {\em et~al.},
  {{\tt arXiv:0905.2160 [hep-lat]}}.

\bibitem{NaLat09}
H.~Na,
\newblock private communication and talk presented at this conference.

\bibitem{Maiani:1990ca}
L.~Maiani and M.~Testa, {\em Phys. Lett.} {\bf B245} (1990) 585.

\bibitem{Lellouch:2000pv}
L.~Lellouch and M.~Luscher, {\em Commun. Math. Phys.} {\bf 219} (2001) 31.

\bibitem{Kim:2002np}
C.-h. Kim and N.~H. Christ, {\em Nucl. Phys. Proc. Suppl.} {\bf 119} (2003)
  365.

\bibitem{Lin:2003tn}
C.~J.~D. Lin, G.~Martinelli, E.~Pallante, C.~T. Sachrajda, and G.~Villadoro,
  {\em Phys. Lett.} {\bf B581} (2004) 207.

\bibitem{Christ:2005gi}
N.~H. Christ, C.~Kim, and T.~Yamazaki, {\em Phys. Rev.} {\bf D72} (2005) 114506.

\bibitem{Kim:2008dr}
C.~Kim, {\em Phys. Rev.} {\bf D80} (2009) 014505.

\bibitem{LightmanLat09}
M.~Lightman,
\newblock talk presented at this conference.

\bibitem{ChristLat09}
N.~Christ,
\newblock talk presented at this conference.

\bibitem{AubinSoni}
C.~Aubin and A.~Soni,
\newblock in preparation.

\bibitem{Buras:1991jm}
A.~J. Buras, M.~Jamin, M.~E. Lautenbacher, and P.~H. Weisz, {\em Nucl. Phys.}
  {\bf B370} (1992) 69.

\bibitem{Gronau:1990ra}
M.~Gronau and D.~London, {\em Phys. Lett.} {\bf B253} (1991) 483.

\bibitem{Gronau:1994bn}
M.~Gronau, J.~L. Rosner, and D.~London, {\em Phys. Rev. Lett.} {\bf 73} (1994) 21.

\bibitem{Bernard:1985wf}
C.~W. Bernard, T.~Draper, A.~Soni, H.~D. Politzer, and M.~B. Wise, {\em Phys.
  Rev.} {\bf D32} (1985) 2343.

\bibitem{Manohar:2000dt}
A.~V. Manohar and M.~B. Wise, {\em Camb. Monogr. Part. Phys. Nucl. Phys.
  Cosmol.} {\bf 10} (2000).

\bibitem{Boyd:1994pa}
C.~G. Boyd and B.~Grinstein, {\em Nucl. Phys.} {\bf B442} (1995) 205.

\end{thebibliography}\endgroup

\end{document}